\documentclass[submission,copyright,creativecommons]{eptcs}
\providecommand{\event}{ITRS 2012}

\usepackage{todonotes}
\usepackage{amsmath}
\usepackage{amsfonts}

\usepackage{float}
\floatstyle{boxed} 
\restylefloat{figure}
\usepackage{subfigure}
\input xypic

\newcommand{\infr}[3]{\displaystyle \frac{{#1}}{{#2}} \mbox{({#3})}}

\newcommand{\bcl}{{\sc bcl}$_{0}$}
\def\textcap{\textstyle\bigcap}
\def\TT{\mathbb{T}}
\def\VV{\mathbb{V}}
\newcommand{\pcmd}[1]{\mbox{\sc #1}}

\xdefinecolor{darkgreen}{rgb}{0,0.5,0}
\xdefinecolor{darkred}{rgb}{0.5,0,0}
\newcommand{\dg}[1]{\color{darkgreen}{#1}}

\newcommand{\pgt}[1]{\mathtt {#1}}
\newcommand{\cbt}[1]{{\dg{\mathtt {#1}}}}
\newcommand{\stype}[1]{\mathit{\color{blue} {#1}}}
\def\scap{\;{\color{blue}\cap}\;}
\newcommand{\svar}[1]{{\color{red}{#1}}}
\def\pR{\pgt{R}}

\title{Using Inhabitation in Bounded Combinatory Logic with Intersection Types for Composition Synthesis}
\author{Boris D\"udder \quad Oliver Garbe \quad Moritz Martens \quad Jakob Rehof
\institute{Faculty of Computer Science\\
Technical University of Dortmund\\
Dortmund, Germany}
\email{\{boris.duedder, oliver.garbe, moritz.martens, jakob.rehof\}@cs.tu-dortmund.de}
\and
Pawe\l\ Urzyczyn
\institute{Institute of Informatics\\
University of Warsaw\\
Warsaw, Poland}
\email{urzy@mimuw.edu.pl}
}
\def\titlerunning{Using Inhabitation for Synthesis}
\def\authorrunning{B. D\"udder, O. Garbe, M. Martens, J. Rehof \& P. Urzyczyn}
\begin{document}
\maketitle

\begin{abstract}
  We describe ongoing work on a framework for automatic composition synthesis from a repository of software components. This work is based on combinatory logic with intersection types. The idea is that components are modeled as typed combinators, and an algorithm for inhabitation --- is there a combinatory term $e$ with type $\tau$ relative to an environment $\Gamma$? --- can be used to synthesize compositions. Here, $\Gamma$ represents the repository in the form of typed combinators, $\tau$ specifies the synthesis goal, and $e$ is the synthesized program. We illustrate our approach by examples, including an application to synthesis from GUI-components.
\end{abstract}

\section{Introduction}
In this paper we describe ongoing work to construct and apply a framework for automatic composition synthesis from software
repositories, based on inhabitation in combinatory logic with intersection types. We describe the basic idea of type-based
synthesis using bounded combinatory logic with intersection types and illustrate an application of the framework to the synthesis of graphical user interfaces (GUIs).
Although our framework is under development and hence results in applications to synthesis are still preliminary, we hope
to illustrate an interesting new approach to type-based synthesis from component repositories. 

In a recent series of papers \cite{RehofU11,RehofU12,RehofEtAlTR12} we have laid the theoretical foundations
for understanding algorithmics and complexity of decidable inhabitation in subsystems of the intersection type system
\cite{code80}. 
In contrast to standard combinatory logic where a fixed basis of combinators is usually considered,
the inhabitation problem considered here is {\em relativized} to an arbitrary environment $\Gamma$ given as part of the input.
This problem is undecidable for combinatory logic, even in simple types, see \cite{RehofEtAlTR12}.
We have introduced finite and bounded combinatory logic
with intersection types in \cite{RehofU11,RehofEtAlTR12} as a possible foundation for type-based composition synthesis. 
{\em Finite combinatory logic} (abbreviated {\sc fcl}) \cite{RehofU11} arises from combinatory logic by restricting combinator types to be monomorphic, and 
{\em $k$-bounded combinatory logic} (abbreviated $\mbox{\sc bcl}_k$) \cite{RehofEtAlTR12} is obtained by imposing the bound $k$ on the depth of types that can be
used to instantiate polymorphic combinator types. It was shown that relativized inhabitation in finite combinatory logic is {\sc Exptime}-complete
\cite{RehofU11}, and that $k$-bounded combinatory logic forms an infinite hierarchy depending on $k$, inhabitation 
being $(k+2)$-{\sc Exptime}-complete for each $k \ge 0$. In this paper, we stay within the lowest level of the hierarchy, $\mbox{\sc bcl}_0$.
We note that, already at this level, we have a framework for \mbox{{\sc 2-Exptime}}-complete synthesis problems, equivalent in complexity
to other known synthesis frameworks (e.g., variants of temporal logic and propositional dynamic logic).

In positing bounded combinatory logic as a foundation for composition synthesis, we consider the {\em inhabitation problem}: Given an environment
$\Gamma$ of typed combinators and a type $\tau$, does there exist a~combinatory term $e$ such that $\Gamma\ \vdash\ e : \tau$?  
For applications in synthesis, we
consider $\Gamma$ as a repository of components represented only by their names (combinators) and their types (intersection types), and~$\tau$ is seen as the specification of a synthesis goal. An inhabitant $e$ is a program obtained by applicative combination of components
in $\Gamma$. The inhabitant $e$ is automatically constructed (synthesized) by the inhabitation algorithm. For applications to synthesis, where the
repository $\Gamma$ may vary, the relativized inhabitation problem is the natural model.

\section{Inhabitation in Finite and Bounded Combinatory Logic}\label{sec:types}
We state the necessary notions and definitions for \textit{finite and bounded combinatory logic with intersection types and subtyping}
\cite{RehofU11,RehofEtAlTR12}. We consider \textit{applicative terms} ranged over by $e$, etc.~and defined as
$$e::=x\;|\;(e\;e'),$$
where $x$, $y$ and $z$ range over a denumerable set of {\em variables} also called {\em combinators}. As usual, we take application of terms to be left-associative. Under these premises any applicative term can be uniquely written as $xe_{1}\ldots e_{n}$ for some $n\geq0$. Sometimes we may also write $x(e_{1},\ldots,e_{n})$ instead of $xe_{1}\ldots e_{n}$. \textit{Types}, ranged over by $\tau$, $\sigma$, etc.~are defined by
$$\tau\,::=\,a\,|\,\tau\to\tau\,|\,\tau\cap\tau$$
where $a,b,c,\ldots$ range over \textit{atoms} comprising \textit{type constants} from a finite set $\mathbb{A}\uplus\{\omega\}$ and \textit{type variables} from a disjoint denumerable set $\mathbb{V}$ ranged over by $\alpha,\beta,\gamma,\ldots$ The constant $\omega$ is the top-element with regard to the subtyping relation $\leq$ defined below. We denote the set of all types by $\mathbb{T}$. As usual, intersections are idempotent, commutative, and associative. Notationally, we take the type constructor $\to$ to be right-associative. A type $\tau\cap\sigma$ is called an \textit{intersection type} or \textit{intersection} \cite{pottinger80,code80} and is said to have~$\tau$ and $\sigma$ as \textit{components}. We sometimes write $\bigcap_{i=1}^{n}\tau_{i}$ for an intersection with $n\geq1$ components. Intersection types come with a natural notion of subtyping, as defined in \cite{code80}. The subtyping relation, denoted by~$\leq$, is the least preorder (reflexive and
transitive relation) on $\TT$ satisfying the following conditions:
\begin{gather*}
\sigma\leq\omega,\quad \omega \leq\omega\to\omega,\quad
\sigma\cap\tau \leq\sigma,\quad \sigma\cap\tau \leq\tau,\quad
\sigma\leq\sigma\cap\sigma;\\
(\sigma\to\tau)\cap(\sigma\to\rho)\leq\sigma\to\tau\cap\rho;\\
\mbox{If~}\sigma\leq\sigma'~\mbox{and}~\tau\leq\tau'~\mbox{then~}
\sigma\cap\tau\leq\sigma'\cap\tau'~\mbox{and}~
\sigma'\to\tau\leq\sigma\to\tau'.
\end{gather*}
Subtyping is used in the systems of \cite{RehofU11,RehofEtAlTR12}, and it is decidable in polynomial time \cite{RehofU11}. We say that two types $\tau$ and $\sigma$ are \textit{equivalent} if and only if $\tau\leq\sigma$ and $\sigma\leq\tau$.

If $\tau=\tau_{1}\to\cdots\to\tau_{n}\to\sigma$ we write $\sigma=\mathit{tgt}_{n}(\tau)$ and $\tau_{i}=\mathit{arg}_{i}(\tau)$ for $i\leq n$ and we say that $\sigma$ is a \textit{target} type of $\tau$ and  $\tau_{i}$ are \textit{argument} types of $\tau$. A type of the form $\tau=\tau_{1}\to\cdots\to\tau_{n}\to a$ with $a\neq\omega$ an atom is called a \textit{path} of length $n$. A type is {\em organized} if it is an intersection of paths. For every type $\tau$ there is an equivalent organized type $\bar{\tau}$ that is computable in polynomial time \cite{RehofU12}. Therefore, in the following we assume all types to be organized. For $\sigma\in\mathbb{T}$ we denote by $\mathbb{P}_{n}(\sigma)$ the set of all paths of length greater than or equal to $n$ in $\sigma$ and by $\parallel \! \sigma \!\parallel$ the path length of $\sigma$ which is defined to be the maximal length of a path in $\sigma$. Define the set $\mathbb{T}_{0}$ of {\em level $0$ types} by $\TT_0 = \{\bigcap_{i\in I}a_{i}\ |\ a_{i}\text{ an atom, }I\mbox{ a finite index set}\}$. Thus, level $0$ types comprise of atoms and intersections of such.
We write $\TT_0(\Gamma,\tau)$ to denote the set of level $0$ types with atoms from $\Gamma$ and $\tau$. Note that $\omega$ is also contained in $\TT_0(\Gamma,\tau)$. A \textit{substitution} is a function $S:\mathbb{V}\to\mathbb{T}_{0}$ such that $S$ is the identity everywhere but on a finite subset of $\mathbb{V}$. We tacitly lift $S$ to a function on types, $S:\mathbb{T}\to\mathbb{T}$, by homomorphic extension. A \textit{type environment} $\Gamma$ is a finite set of type assumptions of the form $x:\tau$.

\begin{figure}[h!]
\vspace{2mm}
\[
\begin{array}{lcl}
\infr{[S: \VV \to \TT_0]}
{\Gamma, x: \tau\ \vdash\ x : S(\tau)}
{var} & \ \ \ \ &
\infr{\Gamma\ \vdash\ e : \tau \rightarrow \tau' \ \ \ \Gamma\ \vdash\ e ': \tau}
{\Gamma\ \vdash\ (e \ e') : \tau'}{\mbox{$\to$}E} \\
& & \\
& & \\
\infr{\Gamma\ \vdash\ e : \tau_1 \ \ \ \ \Gamma\ \vdash\ e : \tau_2}
{\Gamma\ \vdash\ e : \tau_1 \cap \tau_2}
{\mbox{$\cap$}I} & \ \ \ \ &
\infr{\Gamma\ \vdash\ e : \tau \ \ \ \ \tau \le \tau'}
{\Gamma\ \vdash\ e : \tau'}{\mbox{$\le$}} 
\end{array}
\]
\caption{$\mbox{\sc bcl}_0(\cap,\le)$}\label{fig:fcl-sub}
\end{figure}

The type rules for {\em $0$-bounded combinatory logic with intersection types and subtyping}, denoted $\mbox{\sc bcl}_0(\cap,\le)$ or simply \bcl, as presented in
\cite{RehofEtAlTR12}, are given in Figure~\ref{fig:fcl-sub}. The bound $0$ is enforced by the fact that only
substitutions $S$ mapping type variables to level $0$ types in $\TT_0$ are allowed in rule (var). In effect, $\mbox{\sc bcl}_0$
allows a limited form of polymorphism of combinators in $\Gamma$, where type variables can be instantiated with atomic types
or intersections of such.
{\em Finite combinatory logic with intersection types and subtyping}, denoted {\sc fcl}$(\cap,\le)$, as presented in \cite{RehofU11}, is the monomorphic restriction of $\mbox{\sc bcl}_0(\cap,\le)$ where
the substitutions $S$ in rule (var) of Figure~\ref{fig:fcl-sub} are required to be the identity. Hence, rule (var) simplifies to the
axiom $\Gamma, x: \tau\ \vdash\ x : \tau$.

We consider the \textit{relativized inhabitation problem}:
\begin{center}
{\em Given an environment $\Gamma$ and a type $\tau$, does there exist an applicative term $e$ such that $\Gamma\ \vdash\ e:\tau$?}
\end{center}
We sometimes write $\Gamma\ \vdash\ ? : \tau$ to indicate an inhabitation goal.
In~\cite{RehofU11} it is shown that deciding inhabitation in $\mbox{\sc fcl}(\cap,\le)$ is {\sc Exptime}-complete. The 
lower bound is by reduction from the intersection non-emptiness problem for finite bottom-up tree automata,
and the upper-bound is by constructing a~polynomial space bounded alternating Turing machine (ATM) \cite{Chandra81}. 
In~\cite{RehofEtAlTR12} it is shown that $k$-bounded combinatory logic (where substitutions are allowed in rule (var) mapping
type variables to types of depth at most $k$) is $(k+2)$-{\sc Exptime}-complete for every $k \ge 0$, and hence the lowest
level of the bounded hierarchy $\mbox{\sc bcl}_0(\cap,\le)$ is $2$-{\sc Exptime}-complete.
The lower bound for $\mbox{\sc bcl}_0$
is by reduction from acceptance of an exponential space bounded ATM. 
\begin{figure}[h!]
\[
\begin{array}{ll}
 & \mathit{Input}: \ \  \Gamma, \tau \\ 
& \\
1  & \ \ \mbox{// loop} \\
2 & \ \ \ \ \ \ \ \mbox{\sc choose} \ (x:\sigma) \in \Gamma ; \\
3 & \ \ \ \ \ \ \  \sigma' := \textcap\{S(\sigma) \mid S: \mathit{Var}(\Gamma,\tau) \to \TT_0(\Gamma,\tau) \}; \\
4 & \ \ \ \ \ \ \ \mbox{\sc choose} \ n \in \{0,\ldots, \parallel \! \sigma' \!\parallel \} ; \\
5 & \ \ \ \ \ \ \ 
\mbox{\sc choose} \ P \subseteq \mathbb{P}_n(\sigma') ; \\
&  \\
6 &  \ \ \ \ \ \ \  
\mbox{\sc if } (\bigcap_{\pi \in P}  \mathit{tgt}_{n}(\pi) \le \tau) \mbox{ {\sc then}} \\
7 & \ \ \ \ \ \ \ \ \ \ \ \mbox{\sc if } (n=0) \mbox{ {\sc then accept}} ;  \\
8 & \ \ \ \ \ \ \ \ \ \ \
\mbox{{\sc else} } \\
9 & \ \ \ \ \ \ \ \ \ \ \ \ \ \ \
\mbox{\sc forall}(i = 1 \ldots n) \\ 
10 &  \ \ \ \ \ \ \ \ \ \ \ \ \ \ \ \ \ \ \ \ \ \ \ \ \ \tau := \bigcap_{\pi \in P}  \mathit{arg}_{i}(\pi) ; \\
11 & \ \ \ \ \ \ \ \ \ \ \ \ \ \ \ \mbox{\sc goto line} \ 2 ; \\
12 & \ \ \ \ \ \ \ \mbox{\sc else reject}; \\
 &  \\
\end{array}
\]
\caption{Alternating Turing machine $\mathcal{M}$ deciding inhabitation 
for $\mbox{\sc bcl}_0(\cap,\le)$}\label{fig:tmb}
\end{figure}

The $2$-{\sc Exptime} (alternating exponential space) algorithm is shown in
Figure \ref{fig:tmb}. In Figure~\ref{fig:tmb} we use shorthand 
notation for ATM-instruction sequences starting from existential states
($\mbox{\sc choose} \ldots$) and instruction sequences starting from
universal states ($\pcmd{forall}(i = 1 \ldots n)\, s_i$).  
A~command of the form $\mbox{\sc choose} \ x \in P$ branches from an
existential state 
to successor states in which $x$ gets assigned distinct elements of $P$.
A command of the form 
$\pcmd{forall}(i = 1 \ldots n) \, s_i$ branches from a universal state
to successor states from
which each instruction sequence $s_i$ is executed. The machine is exponential space bounded, because the set
of substitutions
$\mathit{Var}(\Gamma,\tau) \to \TT_0(\Gamma,\tau)$ is exponentially bounded. We refer to \cite{RehofEtAlTR12} for further
details.

\section{Synthesis from Component Repositories}
In this section we briefly summarize some 
main points of our methodology for composition synthesis, and we illustrate some of the main principles
by an idealized example (the reader might want to take a preliminary look at the example in Section~\ref{subsec:example-repository} first). 
We should emphasize that we only aim at an intuitive presentation of the general idea in broad outline, and
there are many further aspects to our proposed method that cannot be discussed here for space reasons. The paper \cite{BEAT13} contains a more theoretical account of the methodology and has further examples.

\subsection{Basic principles}
\label{subsec:basic-principles}

\subsubsection*{Semantic specification} It is well known that intersection types can be used to
specify deep semantic properties in the $\lambda$-calculus. The system characterizes the strongly normalizing terms 
\cite{pottinger80,code80}, the inhabitation problem is closely related to the $\lambda$-definability
problem \cite{ssalv09,SMGB12}, and our work on bounded combinatory logic
\cite{RehofU11,RehofEtAlTR12} shows that $k$-bounded inhabitation can code any exponential
level of space bounded alternating Turing machines, depending on $k$. Many existing applications
of intersection types testify to their expressive power in various applications. Moreover, it is
simple to prove but interesting to note that we can specify any given term $e$ uniquely: 
there is an environment $\Gamma_e$ and a type $\tau_e$ such that $e$ is the unique
term with $\Gamma_e\ \vdash\ e : \tau_e$ (see \cite{RehofU11}).

\subsubsection*{A type-based, taxonomic approach} It is a possible advantage of the type-based approach
advocated here (in comparison to, e.g., approaches based on temporal logic) that types
can be naturally associated with code, because application programming interfaces (APIs) already have types. In our applications, we
think of intersection types as hosting, in principle, a two-level type system, consisting of {\em native types}
and {\em semantic types}. Native types are types of the implementation language, whereas semantic
types are abstract, application-dependent conceptual structures, drawn, e.g., from a 
taxonomy (domain ontology). For example, we might consider a~specification
\[
\cbt{F} : ((\pgt{real} \times \pgt{real}) \scap \stype{Cart} \to (\pgt{real} \times \pgt{real}) \scap \stype{Pol}) \scap \stype{Iso}
\]
where native types ($\pgt{real}$, $\pgt{real} \times \pgt{real}, \ldots$) 
are qualified, using intersections with semantic types (in the example, $\stype{Cart}, \stype{Pol}, \stype{Iso}$)
expressing (relative to a given conceptual taxonomy) interesting domain-specific properties of the function (combinator) $\cbt{F}$ --- 
e.g., that it is an isomorphism transforming Cartesian to polar coordinates. More generally, we can think of semantic types as
organized in any system of finite-dimensional feature spaces (e.g., $\stype{Cart}, \stype{Pol}$ are features of coordinates,
$\stype{Iso}$ is a feature of functions) whose elements can be mapped onto the native API using intersections, at any level of
the type structure.

\subsubsection*{Level $0$-bounded polymorphism}
The main difference between {\sc fcl} and $\mbox{\sc bcl}_0$ lies in succinctness of $\mbox{\sc bcl}_0$. For example, consider that we can represent any finite function $f: A \to B$ as an intersection type $\tau_f = \textcap_{a \in A} a \to f(a)$, where elements of $A$ and $B$ are type constants. Suppose we have combinators $(F_i : \tau_{f_i})\in\Gamma$, and we want to synthesize compositions of such functions represented as types (in some of our applications they could, for example, be refinement types \cite{frepfe91}). We might want to introduce composition combinators of arbitrary arity, say $g: (A\to A)^n\to (A\to A)$. In the monomorphic system, a function table for $g$ would be exponentially large in $n$. The single declaration $G: (\alpha_0 \to \alpha_1)\to (\alpha_1 \to\alpha_2)\to\cdots\to(\alpha_{n-1}\to\alpha_n)\to (\alpha_0 \to \alpha_n)$ in $\Gamma$ can be used in \bcl\ to represent~$g$. Through level $0$ polymorphism, the action of $g$ is thereby fully specified. 

Generally, the level $\mbox{\sc bcl}_0$ is already very expressive 
(the inhabitation problem for $\mbox{\sc bcl}_0$ is equivalent to the acceptance problem for alternating exponential space bounded Turing machines, hence $2$-{\sc Exptime} complete \cite{RehofEtAlTR12}). The question of expressive power in practice (how easy or hard it is to specify given classes of practical problems) is harder to answer in general and at this stage of our experience. So far, we have found the formalism of intersection types to be very versatile. More experimental work is needed, however. Another question in this context that would be interesting to consider in future work is the connection to temporal logic synthesis problems (many of which are also $2$-{\sc Exptime} complete).

\subsubsection*{Typed repositories as composition logic programs}
When considering the inhabitation problem $\Gamma\ \vdash\ ? : \tau$ as a foundation for synthesis, it may be useful to think of $\Gamma$ as a form of generalized logic program, broadly speaking, along the lines of the idea of
proof theoretical logic programming languages proposed by Miller et al.~\cite{MillerNPS91}. Under this viewpoint, solving the inhabitation problem $\Gamma\ \vdash\ ? : \tau$ means evaluating the program $\Gamma$ against the goal $\tau$: each typed combinator $F : \sigma$ in $\Gamma$ names a single logical ``rule" (type $\sigma$) in an implicational logic, and the repository $\Gamma$ (a collection of such rules) constitutes a logic ``program", which, when given a goal formula~$\tau$ (type inhabitation target), determines the set of solutions (the set of inhabitants). In other words, the ``rule" (type) of a combinator expresses how the combinator composes with other combinators and how its use contributes to goal resolution in the wider ``program" (repository, $\Gamma$). Indeed, we can view the search procedure of the inhabitation algorithm shown in Figure~\ref{fig:tmb} as an operational semantics for such programs.

\subsection{An example repository}
\label{subsec:example-repository}

We consider a simple, idealized example to illustrate some key ideas in synthesis based on bounded combinatory logic. Consider the section of a repository of functions shown in Figure~\ref{fig:exa}, where the native API of a tracking service is given as a type environment consisting of bindings $\cbt{f} : \pgt{T}$ where $\cbt{f}$ is the name of a function (combinator), and $\pgt{T}$ is a native (implementation) type. We can think of the native repository as a Java API, for example, where the native type $\pR$ abbreviates the type $\pgt{real}$.

The intended meaning and use of the repository is as follows. The function $\cbt{Tr}$ can be called with no arguments
and returns a data structure of type $\pgt{D}((\pR,\pR), \pR, \pR)$ which indicates the position of the caller at the time
of call and the temperature at that position and that time. Thus, the function $\cbt{Tr}$ could be used by a moving object to track
itself and its temperature as it moves. The tracking service might be useful in an intelligent logistics application, where
an object (say, a container) keeps track of its own position (coordinates at a given point in time) and condition
(temperature). Thus, the first component of the structure
$\pgt{D}$ (a pair of real numbers) gives the $2$-dimensional Cartesian coordinate of the caller at the time of call,
the second component (a real number) indicates the time of call, and the third component (a~real number) indicates
the temperature.

In addition to the tracking function $\cbt{Tr}$ the repository contains a number of auxiliary functions which can be used
to project different pieces of information from the data structure $\pgt{D}$, with $\cbt{pos}$ returning the position
(coordinate and time), $\cbt{cdn}$ projects the coordinates from the components of a position, $\cbt{fst}$ and $\cbt{snd}$ project components of a coordinate, and $\cbt{tmp}$ projects
the temperature. Finally, there are conversion functions, $\cbt{cc2pl}$ and $\cbt{cl2fh}$, which convert from
Cartesian to polar coordinates and from Celsius to Fahrenheit, respectively.

\begin{figure}[h!]
\[
\begin{array}{lcl}
\cbt{Tr} & : & \pgt{()} \to \pgt{D}((\pR,\pR), \pR, \pR) \\
\cbt{pos} & : & \pgt{D}((\pR,\pR), \pR, \pR) \to ((\pR,\pR), \pR) \\
\cbt{cdn} & : &  ((\pR,\pR), \pR) \to (\pR, \pR) \\
\cbt{fst} & : & (\pR, \pR) \to \pR \\
\cbt{snd} & : & (\pR, \pR) \to \pR \\
\cbt{tmp} & : & \pgt{D}((\pR,\pR), \pR, \pR) \to \pR \\
\cbt{cc2pl} & : & (\pR,\pR) \to (\pR,\pR) \\
\cbt{cl2fh} & : & \pR \to \pR
\end{array}
\]
\caption{Section of repository implementing a tracking service (native API)}\label{fig:exa}
\end{figure}

\begin{figure}[h!]
$$\hfil\spreaddiagramrows{-1.2pc}\spreaddiagramcolumns{-2.pc}
\xymatrix{&&&&&&&&&&&&&&&\stype{\tiny{TrackData}}\ar@{--}[dllllll]\ar@{--}[drrrrrr]&&&&&&&\\
&&&&&&&&&\stype{\tiny{Pos}}\ar@{--}[dlllll]\ar@{--}[drrrrr]&&&&&&&&&&&&\stype{\tiny{Temp}}\ar@{-}[dl] \ar@{-}[dr]&\\
&&&& \stype{\tiny{Coord}}\ar@{-}[dlll]\ar@{-}[drrr] &&&&&&&&&& \stype{\tiny{Time}}\ar@{-}[dl]\ar@{-}[dr] &&&&&&\stype{\tiny{Cel}}&& \stype{\tiny{Fh}} \\
&\stype{\tiny{Cart}}\ar@{--}[dl]\ar@{--}[dr]&&&&&& \stype{\tiny{Polar}}\ar@{--}[dl]\ar@{--}[dr]&&&&&&\stype{\tiny{Gpst}}&&\stype{\tiny{Utc}}&&&&&&&\\
\stype{\tiny{Cx}}&&\stype{\tiny{Cy}}&&&&\stype{\tiny{Radius}}&&\stype{\tiny{Angle}}&&&&&&&&&&&&&&
}$$
\caption{Semantic structures}\label{fig:exb}
\end{figure}
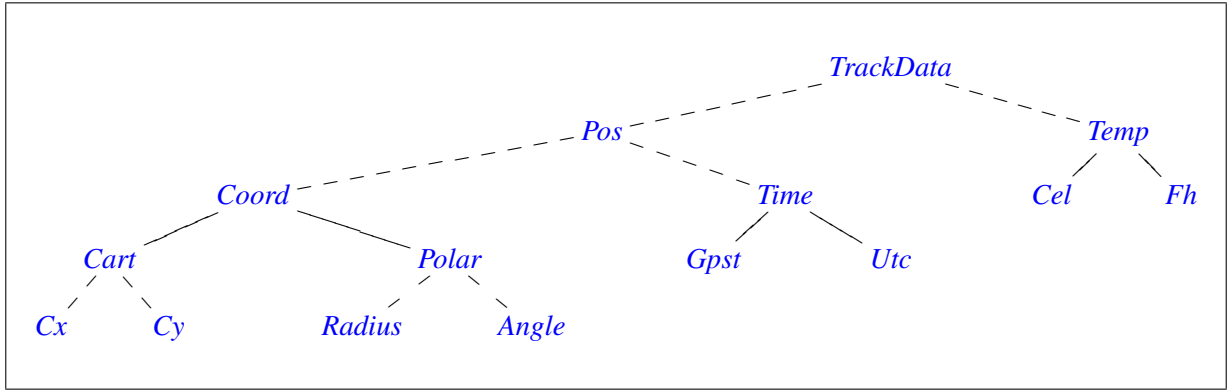

Now, the problem with the standard, native API shown in Figure~\ref{fig:exa} is that it does not express
any of the semantics of its intended use as described above. The basic idea behind combinatory logic synthesis with intersection types is that we
can {\em use intersection types to superimpose conceptual structure onto the native API in order to express semantic
properties}. In order to do so, we must first specify a suitable conceptual structure to capture the intended semantics.
Figure~\ref{fig:exb} shows one such possible structure, which is intended to capture the semantics explained informally
for our example above. The structure is given in the form of a taxonomic tree, the nodes of which are {\em semantic type names},
and where dotted lines indicate structure containment
(for example, elements of the semantic type $\stype{TrackData}$ contain elements of semantic type $\stype{Pos}$ and $\stype{Temp}$), and
solid lines indicate subtyping relationships (for example, $\stype{Cart}$ and $\stype{Polar}$ are subtypes of
$\stype{Coord}$). We are assuming a situation in which certain semantic types can be represented in different ways (as is commonly
the case), e.g., we have $\stype{Time}$ either as GPS Time ($\stype{Gpst}$) or as Universal Time ($\stype{Utc}$), we have
temperature ($\stype{Temp}$) either in Celsius ($\stype{Cel}$) or in Fahrenheit ($\stype{Fh}$), and coordinates can be either polar
or Cartesian.

\begin{figure}[h!]
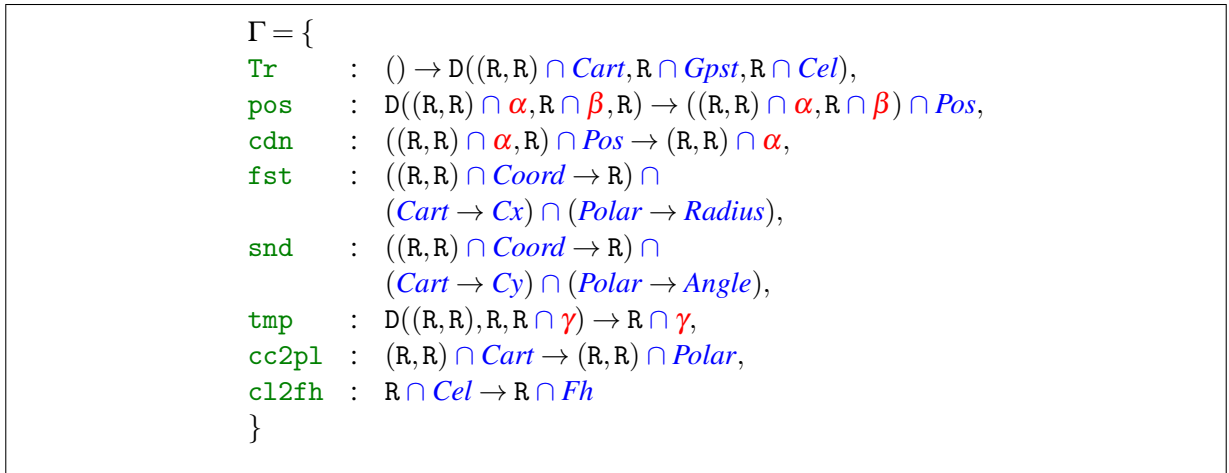

\[
\begin{array}{lcl}
\Gamma = \{ && \\
\cbt{Tr} & : & \pgt{()} \to \pgt{D}((\pR,\pR) \scap \stype{Cart}, \pR \scap \stype{Gpst}, \pR \scap \stype{Cel}), \\
\cbt{pos} & : & \pgt{D}((\pR,\pR) \scap \svar{\alpha}, \pR \scap \svar{\beta}, \pR) \to ((\pR,\pR) \scap \svar{\alpha}, \pR \scap \svar{\beta})
\scap \stype{Pos}, \\
\cbt{cdn} & : &  ((\pR,\pR) \scap \svar{\alpha}, \pR) \scap \stype{Pos} \to (\pR, \pR) \scap \svar{\alpha},  \\
\cbt{fst} & : & ((\pR, \pR) \scap \stype{Coord} \to \pR) \scap \\
&&                 (\stype{Cart} \to \stype{Cx}) \scap (\stype{Polar} \to \stype{Radius}), \\
\cbt{snd} & : & ((\pR, \pR) \scap \stype{Coord} \to \pR) \scap \\
&&                 (\stype{Cart} \to \stype{Cy}) \scap (\stype{Polar} \to \stype{Angle}), \\
\cbt{tmp} & : & \pgt{D}((\pR,\pR), \pR, \pR \scap \svar{\gamma}) \to \pR \scap \svar{\gamma}, \\
\cbt{cc2pl} & : & (\pR,\pR) \scap \stype{Cart} \to 
                   (\pR,\pR) \scap \stype{Polar}, \\
\cbt{cl2fh} & : & \pR \scap \stype{Cel} \to \pR \scap \stype{Fh} \\
\} &&
\end{array}
\]
\caption{Repository with semantic specifications}\label{fig:exc}
\end{figure}

In Figure~\ref{fig:exc} we show the repository of Figure~\ref{fig:exa} with semantic types superimposed onto the native API using intersection types. The superposition of semantic information can be considered as an annotation on the native API. As can be seen, the tracking combinator $\cbt{Tr}$ uses a representation in which coordinates are Cartesian, time is GPS, and temperature is Celsius. Level $0$ polymorphic type variables ($\svar{\alpha}, \svar{\beta},\svar{\gamma}$) are used to succinctly capture semantic information flow, e.g., the combinator $\cbt{pos}$ projects a position ($\stype{Pos}$) from a $\pgt{D}$-typed argument while preserving the semantic information attached to the component types (the variable $\svar{\alpha}$ standing for the semantic qualification of the coordinate component, the variable $\svar{\beta}$ for that of the time component). The types should be readily understandable given the previous explanation of the intended meaning of the API. Notice how we use intersection types to refine \cite{frepfe91} semantic types, as for instance in the type of $\cbt{fst}$, where the type $(\stype{Cart} \to \stype{Cx}) \scap (\stype{Polar} \to \stype{Radius})$ refines the action of $\cbt{fst}$ on the semantic type $\stype{Coord}$.

With the semantically enriched API shown in Figure~\ref{fig:exc} considered as a combinatory type environment $\Gamma$ we can now
ask meaningful questions that can be formalized as synthesis (inhabitation) goals. For example, we can ask whether it is possible to synthesize a 
computation of the current radius (i.e., the radial distance from a standard pole at the current position) by considering the inhabitation
question $\Gamma\ \vdash\ ? : \stype{Radius}$. Sending this question to our inhabitation algorithm gives back the (in this case unique)
solution 
\[
\Gamma\ \vdash\ \cbt{fst} \ (\cbt{cc2pl} \ (\cbt{cdn} \ (\cbt{pos} \ \cbt{Tr()}))) \color{black} : \stype{Radius}
\]
We should note that the concept of a ``current radius" is relative to the given repository. If, for instance, the repository were extended with
a further component $\cbt{origin} : (\pR, \pR) \scap \stype{Cart}$, then we should have the additional solution $\cbt{fst} \ (\cbt{cc2pl} \ \cbt{origin})$ to the query above, which is not the current radius but {\em a}~radius. To express the idea of a current radius in
the extended repository one would have to extend the specifications. Generally, it can be shown \cite{RehofU11}
that any given combinatory term can always be specified uniquely, providing suitable specifications in the repository. But in many practical situations
it is to be expected that queries could yield multiple solutions, which might in some situations be of practical value and in others
the opposite. It is therefore of some interest (also at the time of design of a semantic repository)
to be able to discover and present the structure of solution spaces to queries. As shown in \cite{RehofU11} this can be done, decision procedures
for emptiness, uniqueness and finiteness of solutions are given there. Moreover, one can consider compactly representing larger (or infinite)
sets of solutions, as is briefly discussed in \cite{BEAT13}.

Naturally, the expressive power and flexibility of a repository depends on how it is designed and its type structure
axiomatized (``programmed", referring to the logic programming
analogy mentioned above), and we do not anticipate that our methodology will be applicable to repositories that have not been designed accordingly.

\section{Applications to GUI synthesis}
In this section we focus on the application of inhabitation in bounded combinatory logic in a larger software engineering context. We illustrate how the inhabitation algorithm is integrated into a framework for synthesis from a repository. We have applied the framework to various application domains, including synthesis of control instructions for Lego NXT robots, concurrent workflow synthesis, protocol-based program synthesis, and graphical user interfaces. Below, we will discuss the two last mentioned applications in more detail. 
\subsection{Protocol-Based Synthesis for Windowing Systems}\label{sec:windowing}
Based on protocols we use inhabitation to synthesize programs where the protocols determine the intended program behavior. We give a proof-of-concept example. It illustrates how intersection types can be used to connect different types --- native types and semantic types --- such that data constraints are satisfied whereas semantic types are used to control the result of the synthesis. Figure \ref{fig:protocol} shows a~type environment $\Gamma$ which models a GUI programming scenario for an abstract windowing system. Further, we define the subtyping relations $\pgt{layoutDesktop} \leq \pgt{layoutObj}$ and $\pgt{layoutPDA} \leq \pgt{layoutObj}$. In this scenario we aim to synthesize a program which opens a window, populates it with GUI controls, allows a user interaction, and closes it. The typical data types like $\pgt{wndHnd}$ (window handle) model API data types. Semantic types like $\stype{initialized}$ express the current state of the protocol. Type inhabitation can now be used to synthesize the program described above by asking the inhabitation question $\Gamma\ \vdash\ ? : \stype{closed}$. The inhabitants
$$e_{1}:=\cbt{closeWindow}(\cbt{interact}(\cbt{createControls}(\cbt{openWindow}(\cbt{init}),\cbt{layoutDesktopPC})))$$
$$e_{2}:=\cbt{closeWindow}(\cbt{interact}(\cbt{createControls}(\cbt{openWindow}(\cbt{init}),\cbt{layoutPDAPhone})))$$
share the same type $\stype{closed}$ because both $\pgt{layoutDesktop}$ and $\pgt{layoutPDA}$ are subtypes of $\pgt{layoutObj}$. 
Both terms $e_{1}$ and $e_{2}$ can be interpreted or compiled to realize the intended behavior. These terms are type correct and in addition semantically correct (cf.~Haack et al.~\cite{Wells02}), because all specification axioms defined by the semantic types are satisfied.

\begin{figure}[ht]
	\begin{center}
	  \begin{tabular}{lrl}
	    $\Gamma=\{$&$\cbt{init}$ : & $\stype{start}$,\\
	    &$\cbt{layoutDesktopPC}$ : & $\pgt{layoutDesktop}$,\\
	    &$\cbt{layoutPDAPhone}$ : & $\pgt{layoutPDA}$,\\
	    &$\cbt{openWindow}$ : & $\stype{start}\rightarrow \pgt{wndHnd} \scap \stype{uninitialized}$,\\
	    &$\cbt{createControls}$ : & $\pgt{wndHnd} \scap \stype{uninitialized}\rightarrow \pgt{layoutObj}\rightarrow \pgt{wndHnd} \scap \stype{initialized}$,\\
	    &$\cbt{interact}$ : & $\pgt{wndHnd} \scap \stype{initialized}\rightarrow \pgt{wndHnd} \scap \stype{finished}$,\\
	    &$\cbt{closeWindow}$ : & $\pgt{wndHnd}\scap \stype{finished} \rightarrow \stype{closed}$\}\\
	  \end{tabular}
	\end{center}
	\caption{Type environment $\Gamma$ for protocol-based synthesis in abstract windowing system}\label{fig:protocol}
	\end{figure}

\subsection{GUI Synthesis from a Repository}\label{sec:compmodel}
\begin{figure}[ht]
	\centering
	\subfigure[$ViewIngr\&Prep.ain$]{\includegraphics[scale=0.28]{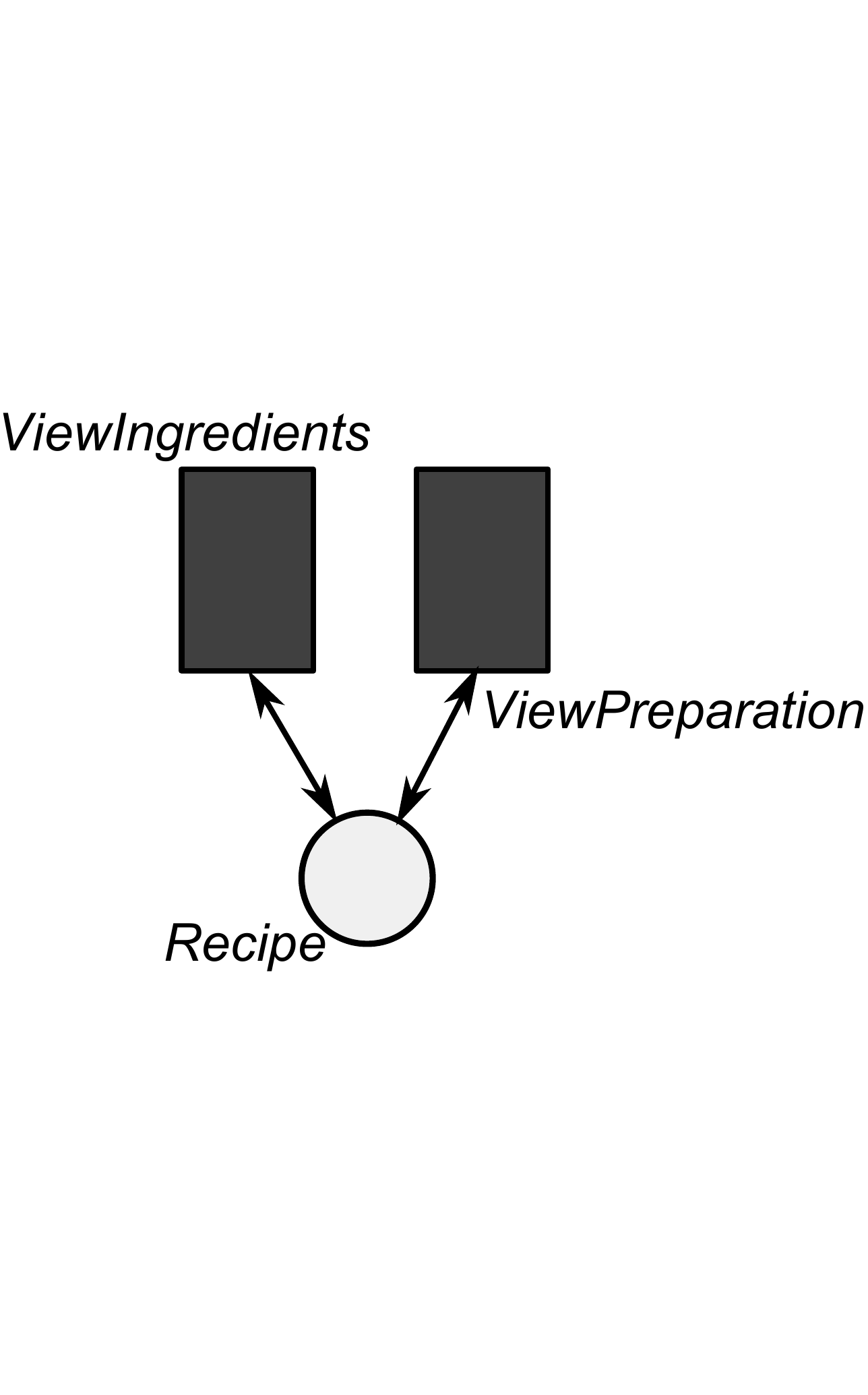}\label{fig:viewIng}}\hspace{.2cm}
	\subfigure[$ViewMeal.ain$]{\includegraphics[scale=0.28]{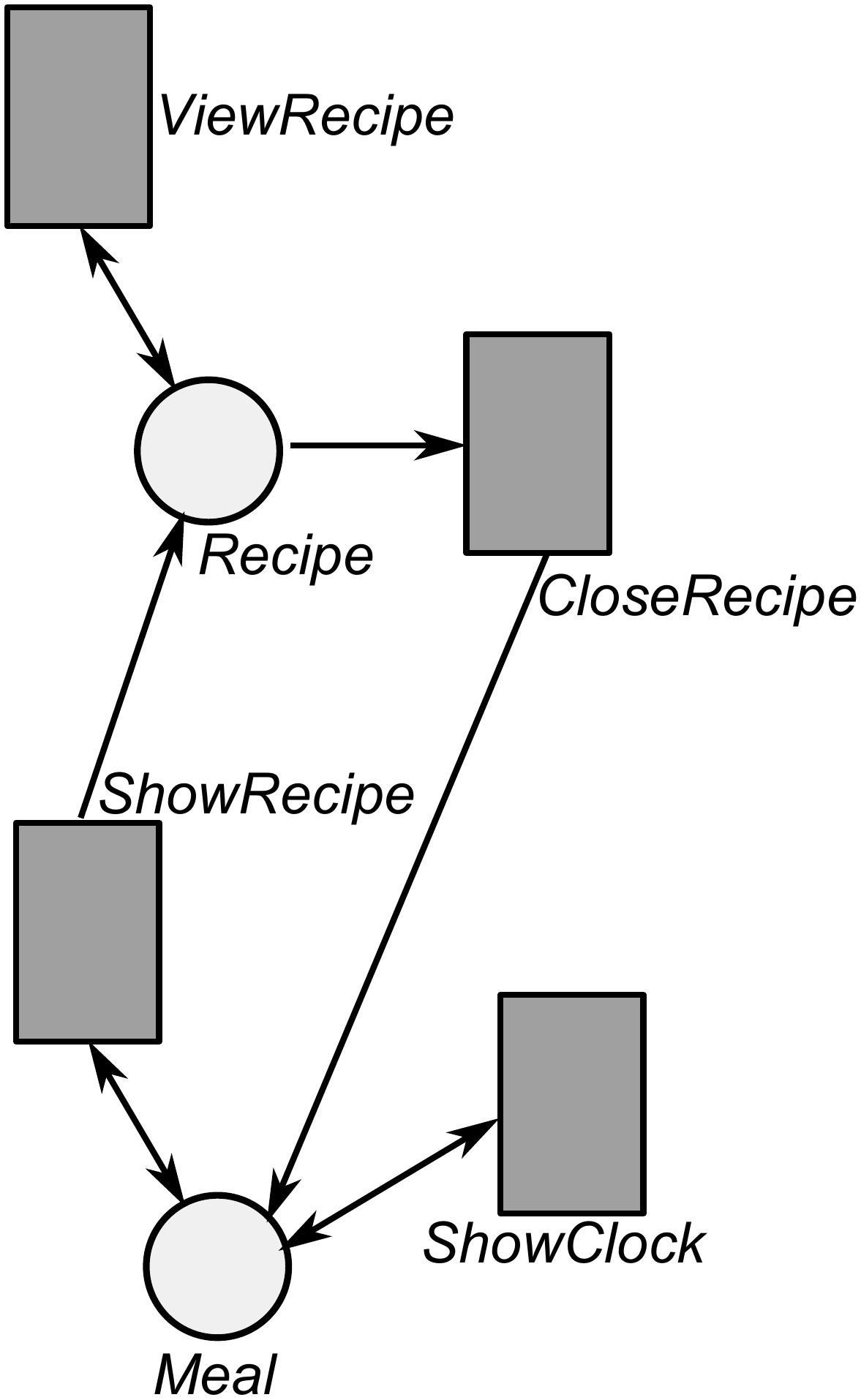}\label{fig:viewMeal}}\hspace{.2cm}
	\subfigure[Resolved GUIF-AIN]{\includegraphics[scale=0.28]{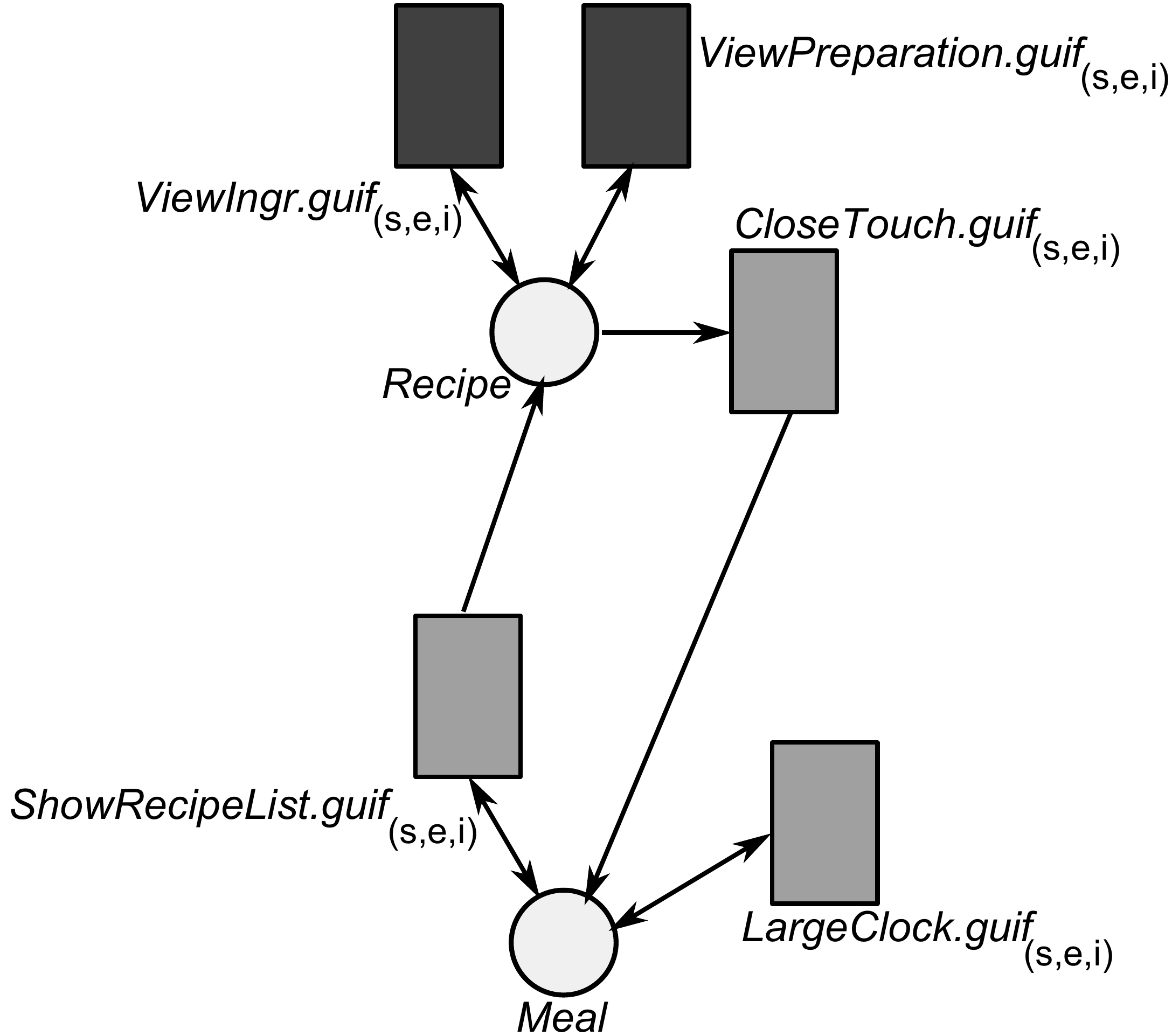}\label{fig:guifAin}}
	\caption{Example AINs}
	\label{fig:example}
\end{figure}

We describe the application of our inhabitation algorithm in a larger framework for component-based GUI-development \cite{Koenigsmann11}, thereby enabling automatic synthesis of GUI-applications from a repository of components. The main point we wish to illustrate is the integration of inhabitation in a more complex software synthesis framework, where combinators may represent a variety of objects, including code templates or abstract structures representing GUI components, into which other components need to be substituted in order to build the desired software application. The main formalism used by \cite{Koenigsmann11} to describe the GUI to be synthesized are abstract interaction nets (AINs). An AIN is an extended Petri net and is used to describe complex interaction patterns occurring in the GUI. The places of an AIN represent the objects of the GUI that are involved in the pattern described. The presence of tokens in a place represents the fact that the corresponding GUI-object is active. The transitions of an AIN represent interactions occurring in the pattern. The AIN itself then describes the pattern as follows: It fixes which objects must be active for an interaction to be possible. By moving tokens it activates and deactivates the objects involved in the interactions and thus describes their effect on these objects. As usual, we graphically depict places as round nodes whereas transitions are depicted by rectangular nodes (cf.~Fig.~\ref{fig:example}).

In our framework GUIs are generated by synthesizing an abstract description of a GUI from a repository of basic GUI building blocks. These blocks are given by GUI-fragments (GUIFs) and AINs. A~GUIF is a single component that has a defined functionality and realizes a certain interaction. GUIFs describe reusable parts of a GUI, for example a drop-down menu. An AIN is a complex building block. It is a structural template which fixes the available objects but regards the transitions as placeholders. To obtain a description of a GUI each placeholder has to be substituted by a GUIF, that realizes the corresponding interaction directly, or (recursively) by an AIN\footnote{The substitution of a transition by an AIN has to obey certain rules.}, that describes a complex interaction pattern that realizes the interaction. The second case may arise, if on a more fine grained level of abstraction, for example, an interaction ``search'' has to be described by an AIN with two transitions which first requires the input of a search term followed by a selection from the listed search results. Since the building blocks may have to adhere to certain constraints each GUIF is linked to a usage context vector describing the contexts the GUIF is suitable for, i.e., the constraints that have to be satisfied if the GUIF is to be used. These context vectors can be use to synthesize a GUI-description which is optimized for certain given constraints. The repository has a hierarchical structure: The interactions used to label the transitions are divided into abstract interactions, alternatives, and variants. Each abstract interaction $i$ has a set of alternatives each of which realizes~$i$. Alternatives can be used to differentiate between various methods to realize an abstract interaction, that have the same result but follow different approaches. For example, an abstract interaction for a search could have two alternatives: the first alternative describing a search by categories, the second alternative describing a search which uses a search mask. Each alternative $a$ has a set of variants each of which realizes $a$. Thus, the alternatives can be understood to logically group variants realizing the same abstract interaction. Each variant $v$ is directly realized by a set of GUIFs or by an AIN that describes a more complex interaction pattern realizing $v$ in the sense described above.

We consider a repository (Figure \ref{fig:catalogueExample}) from a medical scenario \cite{Koenigsmann11, adipositas} for synthesizing GUIs for web applications that support patients to keep diet after medical treatment, for example, by helping plan a~meal. The repository's hierarchical structure is represented by layered lists containing the \textbf{O}bjects, abstract \textbf{I}nteractions, \textbf{A}lternatives, and \textbf{V}ariants. Figure~\ref{fig:catalogueExample} only depicts an excerpt of the repository. In particular, there are more than one alternative to every abstract interaction and more than one variant to every alternative which are not shown in the figure. Below the variants are listed the corresponding GUIFs or AINs. The indices of the GUIFs describe the usage contexts. A~GUIF of usage context $(s,e,i)$, for example, is suitable for a \textbf{s}martphone used by \textbf{e}lderly people with \textbf{i}mpaired vision, whereas GUIFs of usage context $p$ are suitable for desktop \textbf{P}Cs. The AINs $ViewIngr\&Prep.ain$ and $ViewMeal.ain$ are given in Figures \ref{fig:viewIng}, respectively \ref{fig:viewMeal}.
\begin{figure}
	\begin{minipage}[t]{0.495\textwidth}
		\begin{small}
		\begin{description}
			\item[O] Meal
			\begin{description}
				\item[I] Show Clock
				\begin{description}
					\item[A] Show Clock$_{A1}$
					\begin{description}
						\item[V] Show Clock$_{V1}$
						\begin{description}
							\item[GUIF] $LargeClock.guif_{(s,e,i)}$
							\item[GUIF] $DesktopClock.guif_{p}$
							\item[AIN] $DateTime.ain$
						\end{description}
					\end{description}
				\end{description}
				\item[I] Show Recipe
				\begin{description}
					\item[A] Show Recipe$_{A1}$
					\begin{description}
						\item[V] Show Recipe$_{V1}$
						\begin{description}
							\item[GUIF] $ShowRecipeList.guif_{(s,e,i)}$
						\end{description}
					\end{description}
				\end{description}
			\end{description}
			\item[O] Meal Plan
			\begin{description}
				\item[I] View Meal
				\begin{description}
					\item[A] View Meal$_{A1}$
					\begin{description}
						\item[V] View Meal$_{V1}$
						\begin{description}
							\item[AIN] $ViewMeal.ain$
						\end{description}
					\end{description}
				\end{description}
				\item[I] Edit Meal
				\begin{description}
					\item[A] Edit Meal$_{A1}$
					\begin{description}
						\item[V] Edit Meal$_{V1}$
						\begin{description}
							\item[AIN] $EditMeal.ain$
						\end{description}
					\end{description}
				\end{description}
			\end{description}
		\end{description}
		\end{small}
	\end{minipage}
	\begin{minipage}[t]{0.505\textwidth}
		\begin{small}
		\begin{description}
			\item[O] Recipe
			\begin{description}
				\item[I] View Recipe
				\begin{description}
					\item[A] View Recipe Details
					\begin{description}
						\item[V] View Ingr \& Prep
						\begin{description}
							\item[AIN] $ViewIngr\&Prep.ain$
						\end{description}
					\end{description}
				\end{description}
				\item[I] Close Recipe
				\begin{description}
					\item[A] Close Recipe$_{A1}$
					\begin{description}
						\item[V] Close Recipe$_{V1}$
						\begin{description}
							\item[GUIF] $CloseTouch.guif_{(s,e,i)}$
							\item[GUIF] $CloseMouse.guif_{p}$
						\end{description}
					\end{description}
				\end{description}
				\item[I] View Ingredients
				\begin{description}
					\item[A] View Ingredients$_{A1}$
					\begin{description}
						\item[V] View Ingredients$_{V1}$
						\begin{description}
							\item[GUIF] $ViewIngr.guif_{(s,e,i)}$
							\item[GUIF] $IngrDetails.guif_{p}$
						\end{description}
					\end{description}
				\end{description}
				\item[I] View Preparation
				\begin{description}
					\item[A] View Preparation$_{A1}$
					\begin{description}
						\item[V] View Preparation$_{V1}$
						\begin{description}
							\item[GUIF] $ViewPreparation.guif_{(s,e,i)}$
							\item[GUIF] $AnimPreparation.guif_{p}$
						\end{description}
					\end{description}
				\end{description}
			\end{description}
		\end{description}			
		\end{small}
	\end{minipage}
	\caption{Excerpt of repository: Planning a meal}\label{fig:catalogueExample}
\end{figure}

The synthesis problem is now defined as follows: From a given AIN and a usage context vector we want to generate an abstract GUI-description that is optimized for the usage context vector by substituting each transition of the AIN by a suitable GUIF or AIN. Thus, given an AIN, we have to realize each of its transitions \textit{separately} for the given usage context vector. The resulting adapted AIN where all its transitions (in particular also those transitions that emerged due to substitution by further AINs) are realized by GUIFs is called a GUIF-AIN. The transitions of the AIN are realized by means of a recursive algorithm: If a transition labeled $i$ can directly be realized by a GUIF then it is substituted by the GUIF which must be executed whenever the transition is fired. Otherwise, if there is an AIN realizing $i$, then $i$ is substituted by the AIN, whose transitions then have to be recursively realized. Applying this procedure in order to realize $ViewMeal.ain$, results in the GUIF-AIN depicted in Figure~\ref{fig:guifAin}. The transition labeled $ViewRecipe$ of $ViewMeal.ain$ first has to be replaced by $ViewIngr\&Prep.ain$ whose transitions can be directly realized by GUIFs.

In the following we explain how this approach to synthesizing GUIs can be mapped to inhabitation questions such that from the inhabitants a GUIF-AIN realizing the synthesis goal can be assembled: For each abstract interaction, alternative, and variant, as well as for each usage context, we introduce a fresh type constant. The hierarchical structure of abstract interactions, alternatives, and variants is represented by subtyping. We extend $\leq$ with the following additional conditions: We set $a\leq i$ for each abstract interaction $i$ and each of its corresponding alternatives $a$, and we set $v\leq a'$ for each alternative $a'$ and each of its corresponding variants~$v$. We use intersections to represent the usage context vectors: Let $C$ be the set of all usage contexts\footnote{We assume that there are only finitely many usage contexts.}. A usage context given by the non-empty subset $\mathcal{C}\subseteq C$ is in principle represented by the intersection $\bigcap_{c\in\mathcal{C}}c$. However, to prevent name clashing we introduce the type constructor $\mathbf{uc(\cdot)}$ to encapsulate an intersection representing a usage context vector. We assume that $\mathbf{uc}$ is distributive with respect to intersection, i.e., $\mathbf{uc}(c\cap c')=\mathbf{uc}(c)\cap\mathbf{uc}(c')$. The repository of GUI building blocks described above is transformed into a type environment $\Gamma$ as follows. Each GUIF $x$ directly realizes the variant $v_{x}$ it is a child of. Therefore, $x$ must at least be given the type $v_{x}$. Because $x$ further has a usage context vector $\mathcal{C}_{x}$ we augment its type by $\mathbf{uc}(\bigcap_{c\in\mathcal{C}_{x}}c)$, i.e., we get $x:v_{x}\cap\mathbf{uc}(\bigcap_{c\in\mathcal{C}_{x}}c)\in\Gamma$. An AIN $f$ realizes the variant $v_{f}$ it is a child of if all transitions of $f$ are realized. Thus, if $f$ includes $n$ transitions labeled $i_{1}, \ldots, i_{n}$ then it must at least be given the type $i_{1}\to\cdots\to i_{n}\to v_{f}$. Then, inhabiting $v_{f}$ using the combinator $f$ forces \textit{all} arguments of $f$ also to be inhabited in accordance with the fact that the AIN $f$ is realized if all its transitions are realized. We still have to explain how usage context vectors are passed to the arguments of the function type representing the AIN. Consider a fixed context $\mathcal{C}\subseteq C$. The coding $f\,:\,i_{1}\cap\mathbf{uc}(\bigcap_{c\in\mathcal{C}}c)\to\cdots\to i_{n}\cap\mathbf{uc}(\bigcap_{c\in\mathcal{C}}c)\to v_{f}\cap\mathbf{uc}(\bigcap_{c\in\mathcal{C}}c)$ passes the usage context $\mathcal{C}$ to all arguments. This coding ensures that in order to realize the variant $v_{f}$ supplied with the usage context vector $\mathcal{C}$ by using the combinator $f$ \textit{all} $n$ arguments $i_{1},\ldots,i_{n}$ must also be realized in a way which is optimized for $\mathcal{C}$. This reflects the fact that in order to construct a GUIF-AIN for a given usage context \textit{all} its transitions must be realized according to this usage context. However, we of course do not want to restrict the combinators representing AINs to a single usage context vector. It must be possible to represent an AIN by a combinator that can pass an arbitrary usage context vector to its arguments because we do not know beforehand which usage context a GUI should be synthesized for. With monomorphic types ({\sc fcl}) this would lead to the following coding:
$$f\,:\,(i_{1}\to\cdots\to i_{n}\to v_{f})\cap\bigcap_{c\in C}(\mathbf{uc}(c)\to\cdots\to\mathbf{uc}(c)\to\mathbf{uc}(c))\in\Gamma$$
Using polymorphism ($\mbox{\sc bcl}_0$) allows for a more succinct coding, because we may instantiate variables with an intersection representing exactly the usage context vector needed. Thus, we code the AIN $f$ by the combinator:
$$f\,:\,i_{1}\cap \mathbf{uc}(\alpha) \to\cdots\to i_{n}\cap \mathbf{uc}(\alpha) \to v_{f}\cap \mathbf{uc}(\alpha)\in\Gamma$$
The synthesis goal consisting of an AIN $g$ with $m$ transitions labeled $k_{1},\ldots,k_{m}$ and of a usage context vector $\mathcal{C}_{g}$ is represented by asking $m$ inhabitation questions $\Gamma\ \vdash\ ?:k_{j}\cap\mathbf{uc}(\bigcap_{c\in\mathcal{C}_{g}}c)$, for $1\leq j\leq m$.

Part of the type environment $\Gamma_{OM}$ obtained by applying this translation to the example repository in Figure~\ref{fig:catalogueExample} is shown in Figure~\ref{fig:exampleGamma}. As explained above, the subtype relation is extended for abstract interactions, alternatives, and variants. For example, the relations $\pgt{ViewIngr\&Prep} \leq \pgt{ViewRecipeDetails}$ and $\pgt{ViewRecipeDetails} \leq \pgt{ViewRecipe}$ are derived from the repository. Recall that $C_{OM}=\{\stype{p},\stype{s},\stype{e},\stype{i}\}$ is the set of usage contexts, which here contains usage contexts describing GUIFs that are suitable for desktop PCs, smartphones, elderly people, respectively people with impaired vision. For example, if a GUI to display the ingredients and the preparation instructions for a recipe should be realized for a smart-phone used by elderly users with impaired vision, the combinator $\cbt{ViewIngr\&Prep.ain}$ can be instantiated with the type:
\[
\begin{array}{l}
\pgt{ViewIngredients}\scap \mathbf{uc}(\stype{s}\scap \stype{e}\scap \stype{i}) \to \\
\ \ \ \ \ \ \ \  
\pgt{ViewPreparation} \scap \mathbf{uc}(\stype{s}\scap \stype{e}\scap \stype{i}) \to 
\pgt{ViewIngr\&Prep} \scap \mathbf{uc}(\stype{s}\scap \stype{e} \scap \stype{i})
\end{array}
\]
\begin{figure}
	\begin{center}
	\begin{tabular}{rll}
		$\Gamma_{OM}=\lbrace$&$\cbt{LargeClock.guif}$&$:\,\pgt{ShowClock}_{V1}\scap \mathbf{uc}(\stype{s}\scap \stype{e}\scap \stype{i}),$\\
				     &$\cbt{DesktopClock.guif}$&$:\,\pgt{ShowClock}_{V1}\scap \mathbf{uc}(\stype{p}),$\\
				  &$\cbt{ShowRecipeList.guif}$&$:\,\pgt{ShowRecipe}_{V1}\scap \mathbf{uc}(\stype{s}\scap \stype{e}\scap \stype{i}),$\\
				  &$\cbt{CloseTouch.guif}$&$:\,\pgt{CloseRecipe}_{V1}\scap \mathbf{uc}(\stype{s}\scap \stype{e}\scap \stype{i}),$\\
				  &$\cbt{CloseMouse.guif}$&$:\,\pgt{CloseRecipe}_{V1}\scap \mathbf{uc}(\stype{p}),$\\
				  &$\cbt{ViewIngr.guif}$&$:\,\pgt{ViewIngredients}_{V1}\scap \mathbf{uc}(\stype{s}\scap \stype{e}\scap \stype{i}),$\\
				  &$\cbt{IngrDetails.guif}$&$:\,\pgt{ViewIngredients}_{V1}\scap \mathbf{uc}(\stype{p}),$\\
				  &$\cbt{ViewPreparation.guif}$&$:\,\pgt{ViewPreparation}_{V1}\scap \mathbf{uc}(\stype{s}\scap \stype{e}\scap \stype{i}),$\\
				  &$\cbt{AnimPreparation.guif}$&$:\,\pgt{ViewPreparation}_{V1}\scap \mathbf{uc}(\stype{p}),$\\
					&$\cbt{ViewMeal.ain}$&$:\,\pgt{ShowRecipe}\scap \mathbf{uc}(\svar{\alpha}) \to \pgt{ShowClock}\scap\mathbf{uc}(\svar{\alpha}) \to$ \\
					&&$\ \ \ \ \ \ \ \ \pgt{CloseRecipe}\scap \mathbf{uc}(\svar{\alpha}) \to \pgt{ViewRecipe}\scap \mathbf{uc}(\svar{\alpha}) \to$\\
					&&$\ \ \ \ \ \ \ \ \ \ \ \ \pgt{ViewMeal}_{V1}\scap \mathbf{uc}(\svar{\alpha}),$\\
					&$\cbt{ViewIngr\&Prep.ain}$&$:\, \pgt{ViewIngredients} \scap \mathbf{uc}(\svar{\alpha}) \to$ \\
&& $\ \ \ \ \ \ \ \ \pgt{ViewPreparation}\scap \mathbf{uc}(\svar{\alpha}) \to$ \\
&& $\ \ \ \ \ \ \ \ \ \ \ \ \ \ \ \	\pgt{ViewIngr\&Prep}\scap \mathbf{uc}(\svar{\alpha}),\ldots\rbrace$
	\end{tabular}
	\end{center}
	\caption{Part of $\Gamma_{OM}$}\label{fig:exampleGamma}
\end{figure}

In order to explain how to obtain a GUIF-AIN from an inhabitant consider an inhabitant $e$ of the type $j\cap\mathbf{uc}(\bigcap_{c\in\mathcal{C}_{j}}c)$ for an abstract interaction $j$. The corresponding GUIF or GUIF-AIN can recursively be constructed as follows: If $e=x$ where $x$ is a combinator representing the GUIF $x$ then a transition labeled~$j$ is replaced by $x$. Otherwise, $e$ is of the form $fg_{1}\ldots g_{m}$ where $f$ represents an AIN with $m$ transitions. In this case a transition labeled $j$ is replaced by the GUIF-AIN obtained from recursively replacing the transitions of $f$ by the GUIFs or GUIF-AINs corresponding to the terms $g_{k}$. To realize $\pgt{ViewMeal.ain}$ for the context $\{\stype{s},\stype{e},\stype{i}\}$, for example, we ask the four inhabitation questions:
\[
\begin{array}{l}
\Gamma_{OM}\ \vdash\ ?:\pgt{ShowRecipe}\scap \mathbf{uc}(\stype{s} \scap \stype{e} \scap \stype{i}) \\ 
\Gamma_{OM}\ \vdash\ ?:\pgt{ShowClock} \scap \mathbf{uc}(\stype{s} \scap \stype{e} \scap \stype{i}) \\
\Gamma_{OM}\ \vdash\ ?:\pgt{CloseRecipe} \scap \mathbf{uc}(\stype{s} \scap \stype{e} \scap \stype{i}) \\
\Gamma_{OM}\ \vdash\ ?:\pgt{ViewRecipe} \scap \mathbf{uc}(\stype{s} \scap \stype{e} \scap \stype{i})
\end{array}
\]
Figure \ref{fig:guifAin} depicts the result.

The following section discusses (part of) a realization of this approach towards synthesizing GUIs.
\section{Experiments}\label{sec:implementation}

\begin{figure}[h]
  \centering
  \subfigure[Repository with example AIN]{\includegraphics[scale=0.26]{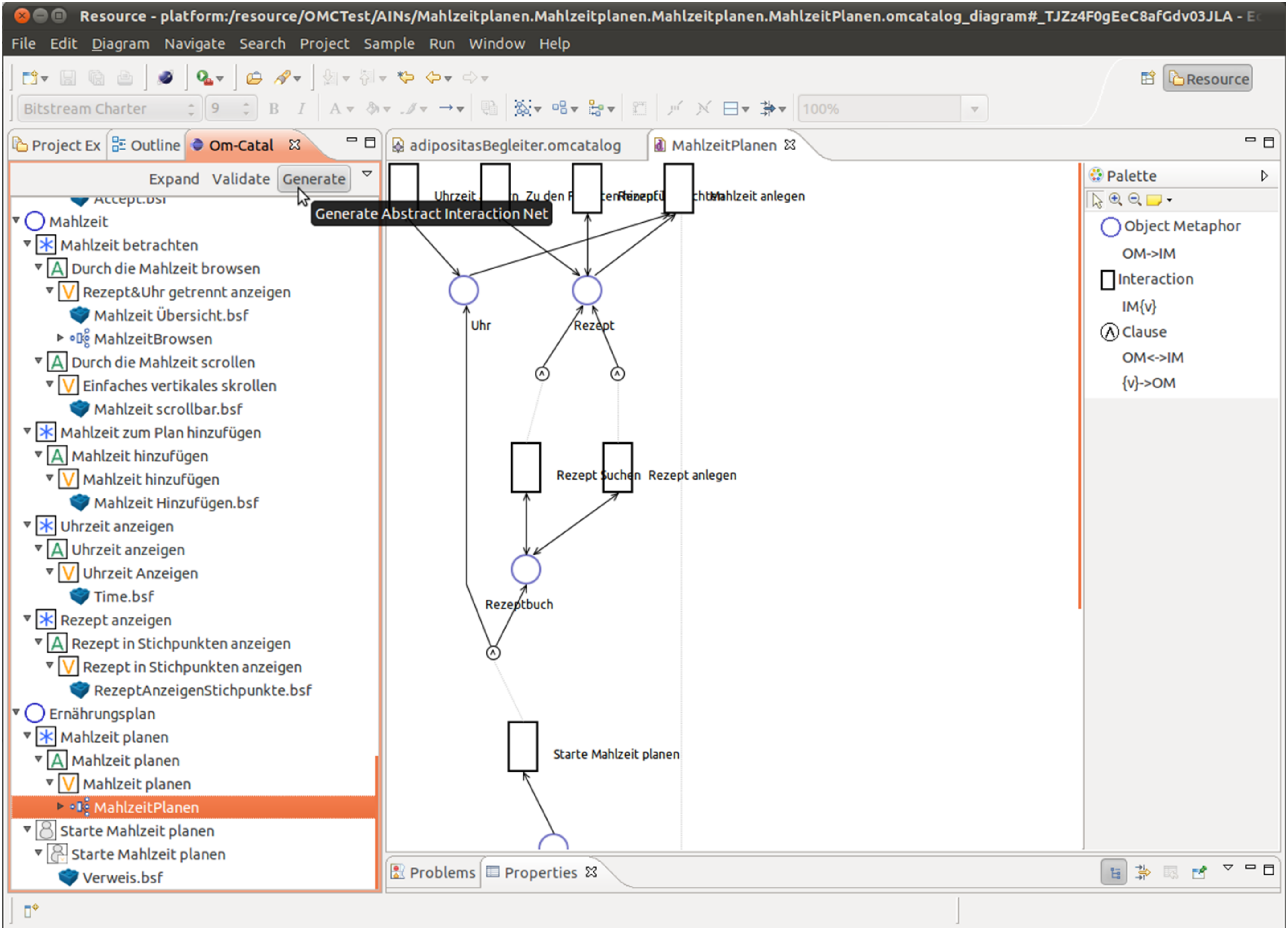}\label{fig:java}}\quad
  \subfigure[Extract of repository]{\includegraphics[scale=0.65]{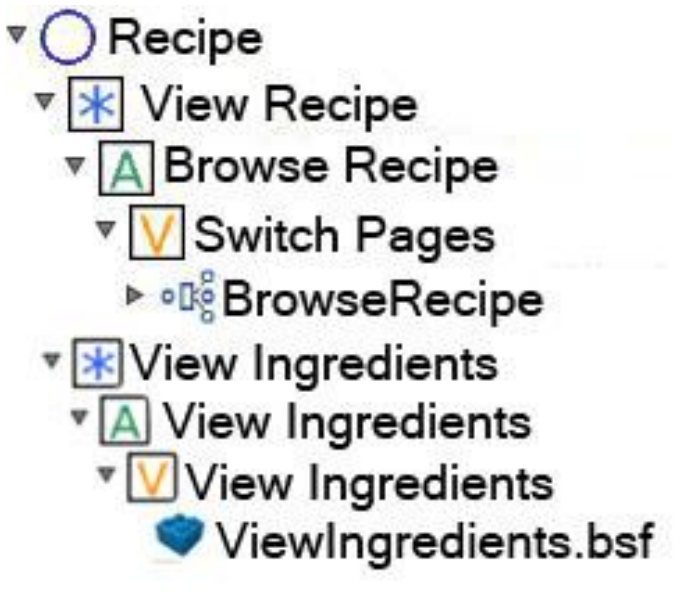}\label{fig:omCatalogue}}
  \caption{GUI for synthesis-framework}
\end{figure}

We presented a prototypical Prolog-implementation of the ATM shown in Figure~\ref{fig:tmb} deciding inhabitation in \bcl\ \cite{TypesDM}. It uses SWI-Prolog \cite{SwiProlog11} and is based on a standard representation of alternation in logic programming \cite{Shapiro84}. This algorithm is used as the core search procedure in a synthesis-framework for GUIs that is based on the coding presented in the previous section. It consists of a Java implementation \cite{Garbe12,Reinke11} based on Eclipse \cite{Eclipse12} providing a suitable data-structure for the repository and a realization of the described translation of the repository into the type environment $\Gamma$. It offers a graphical user interface (Figure \ref{fig:java}) which allows for display and editing of the elements of the repository, posing synthesis-questions, and display and integration into the repository of the results. The constructed inhabitants are directly used to generate corresponding GUIF-AINs from the AINs and GUIFs in the repository. Figure~\ref{fig:omCatalogue} contains an enlarged extract of the repository displayed in Figure \ref{fig:java}. It is a direct manifestation of the repository from which components are drawn for the synthesis of a GUI.

Using this implementation, we conducted some experiments on an extended version of the repository presented in the previous section. The repository contained 12 objects, 27 interactions, 31 alternatives, and 39 variants. There were 9 AINs and 35 GUIFs. In a first prototype the implementation did not treat usage contexts. Instead, a manual post-filtering procedure to identify the solutions best suited for the given usage context was used. Ignoring the usage contexts during inhabitation caused the number of solutions to be very large: Up to 20000 GUIF-AINs were found for some of the AINs in this rather small example. This shows that even for examples of a relatively small size the combinatory explosion may be immense underlining the need for an automation of composition in this case. Furthermore, the sheer number of 20000 solutions made the post-filtering cumbersome if not infeasible. In a second step we incorporated a pre-filtering of $\Gamma$, removing unneeded GUIFs. This resulted in a reduction of the number of suited solutions to approximately 500 which still proved infeasible regarding a composition of the components manually. Including usage contexts by means of intersection types and restricted polymorphism as described in the previous section further reduced the number to only a few solutions. The longest synthesis steps took two to three seconds. 

The presented implementation is only one part in a complete tool chain from design to generation for GUI-synthesis. Here we only focused on the synthesis of an abstract description of the GUI to be generated and its interaction processes. These processes are realized by specifying the necessary GUIFs. Then actual source-code for a web portal server is generated from the synthesized processes by wiring the GUIFs together in a predefined way, thus generating executable GUIs.

\section{Related work}
\label{sec:related}
Our approach is related to adaptation synthesis via proof counting \cite{Wells02,Wells04}, where semantic specifications at the type level
are combined with proof search in a specialized proof system. The idea of adaptation synthesis \cite{Wells02} is closely related to our notion of composition synthesis. However our logic (bounded combinatory logic with intersection types)
is different, and the algorithmic methods are different. In \cite{Wells02} the specification language used is a typed predicate logic,
which is more expressive and more complex than the case of $0$-bounded polymorphism (\bcl), being based on higher-order unification which is undecidable.
The presence of intersection together with $k$-bounded polymorphism yields, of course, enormous
theoretical expressive power (simulation of alternating space bounded Turing machines) and also complexity (nonelementary recursive when the bound $k$ is a parameter). 
A deeper comparison of the relative expressive power in practice must be left for future work. One example, though, which was brought up by one of our reviewers, is worth discussing here, since it points to a methodological point of some generality. The use of predicate logic in \cite{Wells02} allows, e.g., the specification of relations 
between the arguments of a function and its value. For instance, if $P$ is a ternary predicate on reals, we could have the specifications
\[
\cbt{swapArgs} : (\alpha \to \beta \to \gamma) \to \beta \to \alpha \to \gamma \ \mbox{ and } \ \cbt{f}: \pR \to \pR \to \pR
\]
where $\cbt{f}$ satisfies the property $\forall x,y: \pR.\ P(x,y,f x y)$. If we ask for a composition $g$ satisfying the property
$\forall x,y : \pR.\ P(x,y,gyx)$, we get the result $g = \cbt{swapArgs} \ \cbt{f}$. Such properties (as expressed by $P$) can foremostly be expressed
in simple ways in our system when it is possible to name the properties and express their flow through the repository using intersections. In the case
just shown, this can be easily done, by
\[
\cbt{swapArgs} : (\alpha \to \beta \to \gamma) \to ((\beta \to \alpha \to \gamma) \scap \stype{swapped}) \ \mbox{ and } \ \cbt{f}: \pR \to \pR \to \pR
\]
Asking for an inhabitant of $(\pR \to \pR \to \pR) \scap \stype{swapped}$ yields the solution shown. However, it may not always be easy to name
complex relations in such a way that the semantics of them are fully captured. On the other hand, working with polymorphic types and
type constructors in semantic intersection type
components (as in, e.g., a semantic intersection type component $P(\alpha, \beta, \gamma)$) can be enormously expressive. Our lower bound construction in
\cite{RehofEtAlTR12} gives theoretical evidence of this (the lower bound codes the tape of a space bounded Turing machine using such types).  
Further experience is needed to clarify this important class of
questions.

Problems of synthesis from component libraries have been investigated within temporal logic and automata theory
\cite{LustigVardi09}. Our approach is fundamentally different, being based on type theory and combinatory logic, and a direct comparison is therefore precluded. 
However, the fact that both approaches lead to
$2$-{\sc Exptime} complete problems (in our special case of $0$-bounded polymorphism) might suggest that a more detailed comparison could be
an interesting topic for further work. 

\section{Conclusion and further work}
We have introduced the idea of composition synthesis based on combinatory logic with intersection types. Our work is ongoing, and we should emphasize that in the present paper we could only attempt to provide a first encounter with the ideas. There are many avenues for further work. Of foremost importance are optimization of the inhabitation algorithm and further experiments. Although the algorithm matches the worst-case lower bound, there are many interesting principles of optimization to be explored. For better scalability and functionality we have reimplemented the algorithm, using the Microsoft .NET-Framework (F\# and C\#) \cite{dotNet}. This implementation resulted in a tool called \textbf{(CL)S} \cite{cls,BEAT13}. This allowed for a much greater flexibility than the Prolog-based implementation. Some highlights of this implementation are as follows: We parallelized the core procedure of the inhabitation algorithm, allowing for a simultaneous processing of inhabitation questions. This parallelization is an important step towards running our the inhabitation algorithm on multi-CPU architectures --- for example, we deployed a version of this algorithm on a cluster with 1216 cores. Furthermore, we added various features improving usability: The input language is human-readable and closely oriented towards the formal type-language, and there are various graphical output formats, including a display of an execution graph illustrating the tasks processed during inhabitation and of the inhabitants produced. These graphical outputs proved very useful regarding debugging and analysis of the algorithms. We achieved first improvements with regard to optimization of the inhabitation algorithm as discussed above \cite{TR841}. Finally, we added a mechanism which deals with cycles in the inhabitation procedure and thus, allows for a finite representation of infinite sets of inhabitants.
                                                                                                                                                                                                                                                                                                                                                                                                                                                                                                                                                                    
We further pursued the experimental application and evaluation of the ideas described here in a~number of different areas: We used the inhabitation-based synthesis methodology presented here to control synthesis for Lego Mindstorms NXT robots, for example, we were able to synthesize simple pattern-follower programs for these robots with light- or ultrasonic-sensors. Currently, the methodology is applied in a factory planning project which we will report on, soon. Note that \cite{BEAT13} contains a more theoretical exposition of the methodology presented here, also containing further examples. As mentioned above, a comparison to synthesis problems framed in temporal logics could be interesting.

\subsubsection*{Acknowledgements} The authors thank the referees for useful comments.
\bibliographystyle{eptcs}
\bibliography{generic}

\begin{thebibliography}{10}
\providecommand{\bibitemdeclare}[2]{}
\providecommand{\surnamestart}{}
\providecommand{\surnameend}{}
\providecommand{\urlprefix}{Available at }
\providecommand{\url}[1]{\texttt{#1}}
\providecommand{\href}[2]{\texttt{#2}}
\providecommand{\urlalt}[2]{\href{#1}{#2}}
\providecommand{\doi}[1]{doi:\urlalt{http://dx.doi.org/#1}{#1}}
\providecommand{\bibinfo}[2]{#2}

\bibitemdeclare{misc}{cls}
\bibitem{cls}
\emph{\bibinfo{title}{Combinatory Logic Synthesizer --- (CL)S}}.
\newblock
  \bibinfo{howpublished}{\url{http://ls14-www.cs.tu-dortmund.de/index.php/CLS}%
}.

\bibitemdeclare{misc}{dotNet}
\bibitem{dotNet}
\emph{\bibinfo{title}{.NET Development}}.
\newblock
  \bibinfo{howpublished}{\url{http://msdn.microsoft.com/en-us/library/ff361664%
.aspx}}.

\bibitemdeclare{article}{Chandra81}
\bibitem{Chandra81}
\bibinfo{author}{Ashok~K. \surnamestart Chandra\surnameend},
  \bibinfo{author}{Dexter~C. \surnamestart Kozen\surnameend} \&
  \bibinfo{author}{Larry~J. \surnamestart Stockmeyer\surnameend}
  (\bibinfo{year}{1981}): \emph{\bibinfo{title}{Alternation}}.
\newblock {\sl \bibinfo{journal}{J. ACM}} \bibinfo{volume}{28}, pp.
  \bibinfo{pages}{114--133}, \doi{10.1145/322234.322243}.

\bibitemdeclare{article}{code80}
\bibitem{code80}
\bibinfo{author}{Mario \surnamestart Coppo\surnameend} \&
  \bibinfo{author}{Mariangiola \surnamestart Dezani-Ciancaglini.\surnameend}
  (\bibinfo{year}{1980}): \emph{\bibinfo{title}{An Extension of Basic
  Functionality Theory for Lambda-Calculus}}.
\newblock {\sl \bibinfo{journal}{Notre Dame Journal of Formal Logic}}
  \bibinfo{volume}{21}, pp. \bibinfo{pages}{685--693},
  \doi{10.1305/ndjfl/1093883253}.

\bibitemdeclare{misc}{TypesDM}
\bibitem{TypesDM}
\bibinfo{author}{Boris \surnamestart D\"udder\surnameend},
  \bibinfo{author}{Moritz \surnamestart Martens\surnameend} \&
  \bibinfo{author}{Jakob \surnamestart Rehof\surnameend}
  (\bibinfo{year}{2011}): \emph{\bibinfo{title}{Prototype Implementation of an
  Inhabitation Algorithm for FCL$(\cap, \leq)$}}.
\newblock \bibinfo{howpublished}{Presentation at Types 2011 in Bergen, Norway}.

\bibitemdeclare{techreport}{TR841}
\bibitem{TR841}
\bibinfo{author}{Boris \surnamestart D\"udder\surnameend},
  \bibinfo{author}{Moritz \surnamestart Martens\surnameend} \&
  \bibinfo{author}{Jakob \surnamestart Rehof\surnameend}
  (\bibinfo{year}{2012}): \emph{\bibinfo{title}{{Intersection Type Matching and
  Bounded Combinatory Logic (Extended Version)}}}.
\newblock \bibinfo{type}{Technical Report} \bibinfo{number}{841},
  \bibinfo{institution}{Faculty of Computer Science (TU Dortmund)}.
\newblock
  \bibinfo{note}{\url{http://ls14-www.cs.tu-dortmund.de/index.php/Jakob_Rehof_%
Publications\#Technical_Reports}}.

\bibitemdeclare{inproceedings}{RehofEtAlTR12}
\bibitem{RehofEtAlTR12}
\bibinfo{author}{Boris \surnamestart D\"udder\surnameend},
  \bibinfo{author}{Moritz \surnamestart Martens\surnameend},
  \bibinfo{author}{Jakob \surnamestart Rehof\surnameend} \&
  \bibinfo{author}{Pawe\l\ \surnamestart Urzyczyn\surnameend}
  (\bibinfo{year}{2012}): \emph{\bibinfo{title}{{Bounded Combinatory Logic}}}.
\newblock In: {\sl \bibinfo{booktitle}{Computer Science Logic (CSL'12)}}, {\sl
  \bibinfo{series}{LIPIcs}}~\bibinfo{volume}{16},
  \bibinfo{publisher}{Leibniz-Zentrum fuer Informatik}, pp.
  \bibinfo{pages}{243--258}, \doi{10.4230/LIPIcs.CSL.2012.243}.

\bibitemdeclare{unpublished}{Eclipse12}
\bibitem{Eclipse12}
\bibinfo{author}{\surnamestart Eclipse.org\surnameend}:
  \emph{\bibinfo{title}{Eclipse Indigo (3.7) Documentation}}.
\newblock \urlprefix\url{http://help.eclipse.org}.

\bibitemdeclare{inproceedings}{frepfe91}
\bibitem{frepfe91}
\bibinfo{author}{Tim \surnamestart Freeman\surnameend} \&
  \bibinfo{author}{Frank \surnamestart Pfenning\surnameend}
  (\bibinfo{year}{1991}): \emph{\bibinfo{title}{Refinement Types for {ML}}}.
\newblock In: {\sl \bibinfo{booktitle}{ACM Conference on Programming Language
  Design and Implementation (PLDI)}}, \bibinfo{publisher}{ACM}, pp.
  \bibinfo{pages}{268--277}, \doi{10.1145/113445.113468}.

\bibitemdeclare{mastersthesis}{Garbe12}
\bibitem{Garbe12}
\bibinfo{author}{Oliver \surnamestart Garbe\surnameend} (\bibinfo{year}{2012}):
  \emph{\bibinfo{title}{Synthese von Benutzerschnittstellen mit einem
  Typinhabitationsalgorithmus}}.
\newblock \bibinfo{type}{Diploma thesis}, \bibinfo{school}{Technical University
  of Dortmund}.

\bibitemdeclare{inproceedings}{Wells02}
\bibitem{Wells02}
\bibinfo{author}{Christian \surnamestart Haack\surnameend},
  \bibinfo{author}{Brian \surnamestart Howard\surnameend},
  \bibinfo{author}{Allen \surnamestart Stoughton\surnameend} \&
  \bibinfo{author}{Joe~B. \surnamestart Wells\surnameend}
  (\bibinfo{year}{2002}): \emph{\bibinfo{title}{Fully Automatic Adaptation of
  Software Components Based on Semantic Specifications}}.
\newblock In: {\sl \bibinfo{booktitle}{AMAST'02}}, {\sl \bibinfo{series}{LNCS}}
  \bibinfo{volume}{2422}, \bibinfo{publisher}{Springer}, pp.
  \bibinfo{pages}{83--98}, \doi{10.1007/3-540-45719-4\_7}.

\bibitemdeclare{phdthesis}{Koenigsmann11}
\bibitem{Koenigsmann11}
\bibinfo{author}{Thomas \surnamestart K{\"o}nigsmann\surnameend}
  (\bibinfo{year}{2011}): \emph{\bibinfo{title}{Compositional Modelling Ansatz
  zur Benutzerschnittstellengenerierung am Beispiel telemedizinischer
  Anwendungen}}.
\newblock Ph.D. thesis, \bibinfo{school}{Technical University of Dortmund}.

\bibitemdeclare{inproceedings}{adipositas}
\bibitem{adipositas}
\bibinfo{author}{Thomas \surnamestart K\"onigsmann\surnameend} \&
  \bibinfo{author}{Reinholde \surnamestart Kriebel\surnameend}
  (\bibinfo{year}{2008}): \emph{\bibinfo{title}{Digitale Gesundheitsbegleiter
  am Beispiel der Adi\-po\-si\-tas-Nachsorge}}.
\newblock In: {\sl \bibinfo{booktitle}{Proceedings of AAL 2008}},
  \bibinfo{organization}{Verband der Elektrotechnik, Elektronik,
  Informationstechnik}, \bibinfo{publisher}{VDE-Verlag}.

\bibitemdeclare{inproceedings}{LustigVardi09}
\bibitem{LustigVardi09}
\bibinfo{author}{Y.~\surnamestart Lustig\surnameend} \& \bibinfo{author}{M.~Y.
  \surnamestart Vardi\surnameend} (\bibinfo{year}{2009}):
  \emph{\bibinfo{title}{Synthesis from Component Libraries}}.
\newblock In: {\sl \bibinfo{booktitle}{FOSSACS}}, {\sl \bibinfo{series}{LNCS}}
  \bibinfo{volume}{5504}, \bibinfo{publisher}{Springer}, pp.
  \bibinfo{pages}{395--409}, \doi{10.1007/978-3-642-00596-1\_28}.

\bibitemdeclare{article}{MillerNPS91}
\bibitem{MillerNPS91}
\bibinfo{author}{Dale \surnamestart Miller\surnameend},
  \bibinfo{author}{Gopalan \surnamestart Nadathur\surnameend},
  \bibinfo{author}{Frank \surnamestart Pfenning\surnameend} \&
  \bibinfo{author}{Andre \surnamestart Scedrov\surnameend}
  (\bibinfo{year}{1991}): \emph{\bibinfo{title}{Uniform Proofs as a Foundation
  for Logic Programming}}.
\newblock {\sl \bibinfo{journal}{Ann. Pure Appl. Logic}}
  \bibinfo{volume}{51}(\bibinfo{number}{1-2}), pp. \bibinfo{pages}{125--157},
  \doi{10.1016/0168-0072(91)90068-W}.

\bibitemdeclare{incollection}{pottinger80}
\bibitem{pottinger80}
\bibinfo{author}{Garrel \surnamestart Pottinger\surnameend}
  (\bibinfo{year}{1980}): \emph{\bibinfo{title}{A Type Assignment for the
  Strongly Normalizable Lambda-Terms}}.
\newblock In \bibinfo{editor}{J.~\surnamestart Hindley\surnameend} \&
  \bibinfo{editor}{J.~\surnamestart Seldin\surnameend}, editors: {\sl
  \bibinfo{booktitle}{To H. B. Curry: Essays on Combinatory Logic, Lambda
  Calculus and Formalism}}, \bibinfo{publisher}{Academic Press}, pp.
  \bibinfo{pages}{561--577}.

\bibitemdeclare{inproceedings}{BEAT13}
\bibitem{BEAT13}
\bibinfo{author}{Jakob \surnamestart Rehof\surnameend} (\bibinfo{year}{2013}):
  \emph{\bibinfo{title}{{Towards Combinatory Logic Synthesis}}}.
\newblock In: {\sl \bibinfo{booktitle}{Proceedings of BEAT'13}},
  \bibinfo{publisher}{ACM}.

\bibitemdeclare{inproceedings}{RehofU11}
\bibitem{RehofU11}
\bibinfo{author}{Jakob \surnamestart Rehof\surnameend} \&
  \bibinfo{author}{Pawe\l\ \surnamestart Urzyczyn\surnameend}
  (\bibinfo{year}{2011}): \emph{\bibinfo{title}{Finite Combinatory Logic with
  Intersection Types}}.
\newblock In: {\sl \bibinfo{booktitle}{TLCA}}, {\sl \bibinfo{series}{Lecture
  Notes in Computer Science}} \bibinfo{volume}{6690},
  \bibinfo{publisher}{Springer}, pp. \bibinfo{pages}{169--183},
  \doi{10.1007/978-3-642-21691-6\_15}.

\bibitemdeclare{inproceedings}{RehofU12}
\bibitem{RehofU12}
\bibinfo{author}{Jakob \surnamestart Rehof\surnameend} \&
  \bibinfo{author}{Pawe\l\ \surnamestart Urzyczyn\surnameend}
  (\bibinfo{year}{2012}): \emph{\bibinfo{title}{The Complexity of Inhabitation
  with Explicit Intersection}}.
\newblock In \bibinfo{editor}{Robert~L. \surnamestart Constable\surnameend} \&
  \bibinfo{editor}{A.~\surnamestart Silva\surnameend}, editors: {\sl
  \bibinfo{booktitle}{Kozen Festschrift}}, \bibinfo{volume}{LNCS 7230}, pp.
  \bibinfo{pages}{256--270}, \doi{10.1007/978-3-642-29485-3\_16}.

\bibitemdeclare{mastersthesis}{Reinke11}
\bibitem{Reinke11}
\bibinfo{author}{Eugen \surnamestart Reinke\surnameend} (\bibinfo{year}{2011}):
  \emph{\bibinfo{title}{Konzeption und Entwicklung eines Tools zur Modellierung
  von User-Interface-Spezifikationsbausteinen im Rahmen des ``Compositional
  Modeling''-Ansatzes}}.
\newblock \bibinfo{type}{Diploma thesis}, \bibinfo{school}{Technical University
  of Dortmund}.

\bibitemdeclare{inproceedings}{ssalv09}
\bibitem{ssalv09}
\bibinfo{author}{Sylvain \surnamestart Salvati\surnameend}
  (\bibinfo{year}{2009}): \emph{\bibinfo{title}{Recognizability in the Simply
  Typed Lambda-Calculus}}.
\newblock In \bibinfo{editor}{H.~\surnamestart Ono\surnameend},
  \bibinfo{editor}{M.~\surnamestart Kanazawa\surnameend} \&
  \bibinfo{editor}{R.~J. G.~B. \surnamestart de~Queiroz\surnameend}, editors:
  {\sl \bibinfo{booktitle}{WoLLIC}}, {\sl \bibinfo{series}{LNCS}}
  \bibinfo{volume}{5514}, \bibinfo{publisher}{Springer}, pp.
  \bibinfo{pages}{48--60}, \doi{10.1007/978-3-642-02261-6\_5}.

\bibitemdeclare{inproceedings}{SMGB12}
\bibitem{SMGB12}
\bibinfo{author}{Sylvain \surnamestart Salvati\surnameend},
  \bibinfo{author}{Giulio \surnamestart Manzonetto\surnameend},
  \bibinfo{author}{Mai \surnamestart Gehrke\surnameend} \&
  \bibinfo{author}{Henk \surnamestart Barendregt\surnameend}
  (\bibinfo{year}{2012}): \emph{\bibinfo{title}{Loader and Urzyczyn Are
  Logically Related}}.
\newblock In: {\sl \bibinfo{booktitle}{ICALP 12, Automata, Languages, and
  Programming - 39th International Colloquium, Warwick, UK}}, {\sl
  \bibinfo{series}{LNCS}} \bibinfo{volume}{7392},
  \bibinfo{publisher}{Springer}, pp. \bibinfo{pages}{364--376},
  \doi{10.1007/978-3-642-31585-5\_34}.

\bibitemdeclare{article}{Shapiro84}
\bibitem{Shapiro84}
\bibinfo{author}{Ehud~Y. \surnamestart Shapiro\surnameend}
  (\bibinfo{year}{1984}): \emph{\bibinfo{title}{Alternation and the
  Computational Complexity of Logic Programs}}.
\newblock {\sl \bibinfo{journal}{J. Log. Program.}}
  \bibinfo{volume}{1}(\bibinfo{number}{1}), pp. \bibinfo{pages}{19--33},
  \doi{10.1016/0743-1066(84)90021-9}.

\bibitemdeclare{inproceedings}{Wells04}
\bibitem{Wells04}
\bibinfo{author}{J.~B. \surnamestart Wells\surnameend} \&
  \bibinfo{author}{Boris \surnamestart Yakobowski\surnameend}
  (\bibinfo{year}{2005}): \emph{\bibinfo{title}{Graph-Based Proof Counting and
  Enumeration with Applications for Program Fragment Synthesis}}.
\newblock In \bibinfo{editor}{Sandro \surnamestart Etalle\surnameend}, editor:
  {\sl \bibinfo{booktitle}{LOPSTR 2004}}, {\sl \bibinfo{series}{LNCS}}
  \bibinfo{volume}{3573}, \bibinfo{publisher}{Springer}, pp.
  \bibinfo{pages}{262--277}, \doi{10.1007/11506676\_17}.

\bibitemdeclare{article}{SwiProlog11}
\bibitem{SwiProlog11}
\bibinfo{author}{Jan \surnamestart Wielemaker\surnameend}, \bibinfo{author}{Tom
  \surnamestart Schrijvers\surnameend}, \bibinfo{author}{Markus \surnamestart
  Triska\surnameend} \& \bibinfo{author}{Torbj{\"o}rn \surnamestart
  Lager\surnameend} (\bibinfo{year}{2010}): \emph{\bibinfo{title}{SWI-Prolog}}.
\newblock {\sl \bibinfo{journal}{Computing Research Repository}}
  \bibinfo{volume}{abs/1011.5332}.

\end{thebibliography}
\end{document}